\documentclass[12pt]{article}
\usepackage{amsfonts}
\usepackage{amsmath,amssymb}
\usepackage{amsthm}
\usepackage{mathrsfs}
\usepackage{amsxtra}
\usepackage{amstext}
\usepackage{amscd}
\usepackage{latexsym}
\newtheorem{thm}{Theorem}[section]

\newtheorem{lem}[thm]{Lemma}
\newtheorem{prop}[thm]{Proposition}
\newtheorem{defn}[thm]{Definition}

\newtheorem{rem}[thm]{Remark}
\numberwithin{equation}{section}
\def\Tr{\mathrm{Tr}}
\def\tr{\mathrm{tr}}
\def\sgn{\mathrm{sign}}

\def\hB{\hat{B}}

\def\D{{\mathscr D}}
\def\F{{\mathscr F}}
\def\hF{\hat{F}}
\def\hG{\hat{G}}
\def\G{{\mathscr G}}
\def\H{{\mathfrak H}}

\def\L{\hat{\mathcal L}}

\def\hL{\hat{L}}
\def\M{{\mathfrak M}}
\def\P{{\mathcal P}}
\def\p{\hat{\mathscr P}_0} 
\def\R{\mathbb R}

\def\hS{\hat{S}}
\def\C{\mathbb C}

\def\N{\mathbb N}
\def\n{\mathcal N}
\def\sgn{\,\mathrm{sgn}}
\usepackage{color}

\begin{document}
\setcounter{footnote}{0}
\renewcommand{\thefootnote}{\arabic{footnote}}

\begin{center}

\textbf{\large Open Quadratic Fermion Systems and} 

\smallskip

\textbf{\large Algebras of Affine Transformations} 

\bigskip

\large{Hiroshi T}\small{AMURA} \footnote{hiroshi.tamura@komatsu-u.ac.jp}\\

\bigskip

        Komatsu University, \\
        Shichoumachi nu 1-3, Komatsu 923-8511, Japan 

\vspace{1.5cm}

ABSTRACT

\end{center}

We study evolution of open quadratic fermion systems in the  framework of the
quantum Markovian semigroup approach.  
We show that the algebra concerning commutators of Liouvillians for systems of quadratic interacting fermions of finite number, say $\n$, is isomorphic to that of certain affine transformations on the space of square matrices of size $\n$.   
By the use of this algebraic structure, we present a perspective method for solving  
master equations of quadratic  fermion systems.  
Here, we mainly  deal with gauge invariant quadratic interacting fermion systems.  
We briefly mention similar algebraic structures for 
general quadratic fermion systems and quadratic boson systems.        

\bigskip


\section{Introduction}\label{Intro}
The assumption of Markov property in the theory of open quantum systems leads to the quantum master equation \cite{GKS, L, D} as the fundamental equation describing time evolution of density matrices of quantum systems interacting with thermal baths \cite{BP, AL}.     
Several efforts to solve the master equation in the simple and basic cases such as quadratic system of Fermions and Bosons have been made successfully \cite{P}, where   
commutation relations of (super-)operators  acting on density matrices play an important role.   
Moreover, the significance of their Lie algebraic structure has been recognized.  
See \cite{TS} and the references therein.

In this paper, we proceed to study the structure of Lie algebra concerned with commutation relations of operators acting on density matrices to give 
simple representations to solutions for the master equations.  
We consider the models of open quantum systems of quadratic interacting Fermions of an arbitrary but fixed finite number.   
We make a Lie algebra which includes all the Liouvillians of these models with commutator as its Lie bracket and see that it is isomorphic to the Lie algebra of certain group which act as an affine transformation on a space of square matrices of size equal to the number of Fermions.  
Then, we apply it to express evolution of density matrices in perspective formulae.    
We mainly argued the gauge invariant models for open quadratic Fermion systems.   
However, we briefly mention that similar algebraic structures also hold for 
general quadratic Fermion systems as well as quadratic Boson systems.        
As an application, we show how to incorporate the skin effect in Hatano-Nelson model of non-Hermitian quantum mechanics into the formulation of the open quantum systems.  \cite{HN1, HN2} 

This paper is organized as follows.  
In section 2, we introduce the models of open quantum system of gauge invariant quadratic interacting Fermions of an arbitrary but fixed finite number by specifying the form of master equations.  
The commutation relations among Liouvillians of these models are explicitly derived, showing that the set of the Liouvillians forms a Lie algebra.   
In section 3, a group of affine transformations on a space of square matrices of size equal to the number of Fermions is introduced.  
Its Lie algebra is shown to be isomorphic to that of the Liouvillians discussed in section 2.   
We show in Section 4 formulae for the solutions of master equations which is motivated in Section 3, directly.   
Then, by constructing a useful basis of the vector space spanned by density matrices  of the models, asymptotic behavior of the evolution of the density matrices is argued.   
Evolution of Gauss states is considered in connection with the affine transformation discussed in Section 3.     
In section 5, we consider how to take the skin effect known in the non-Hermitian Hatano-Nelson model into Markovian open quantum system.   
Finally, to indicate possible generalizations, we give commutation relations for Liouvillians of the general quadratic interacting fermion systems and Boson systems in Section 6.    

\section{Models}
For  arbitrary but fixed  $\n \in \N$, we consider the $\n$ mode Fock space  $\F$.   
That is, there are linear operators\footnote{Hereafter we use abbreviation ``operator'' for ``linear operator''.} $c_j, c^{\dagger}_j \ (j=1, \cdots, \n) $ acting on $\F$ satisfying the canonical anti-commutation relations:
\begin{equation}
         \{ c_j, c_k \} = 0, \quad  \{ c_j, c^{\dagger}_k \} = \delta_{jk}, \quad  
         \{ c^{\dagger}_j, c^{\dagger}_k \} = 0.  
\label{CAR}
\end{equation}
And there is a specific element $|v \rangle \in \F$ such that 
$ c_j|v \rangle =0 $ holds for all $ j=1, \cdots, \n$  and 
$\{ c_1^{\nu_1}c_2^{\nu_2} \cdots c_{\n}^{\nu_{\n}} |v \rangle \, | \, \nu_j \in \{ 0,1 \} \ (j =1, \cdots,\n) \, \}$ forms a complete orthonormal system of $\F$.  
The identity operator on $\F$ is identified with $ 1 \in \C$. 

Density matrices which represent states of the system of $\n$-fermions are expressed  by positive operators on $\F$ with unit trace.   
By $ \Omega $, we denote the specific density matrix that is  the orthogonal projection to $\C|v \rangle$ (\textit{i.e.}, $\Omega = |v \rangle \langle v |$).  
We introduce the space $\D$ as the Banach algebra of all operators acting on $\F$. 
Note that $\D$ is spanned by the set of all density matrices.  
Since $\F$ is a finite dimensional Hilbert space of dimension $2^{\n}$, $\D$ is also a finite dimensional space of dimension $2^{2\n}$.  
We may consider $\D$ a space of density matrices and a space of observables, simultaneously.

We suppose that the $\n$-Fermion system connects to heat baths and that the density matrix $\rho(t)$ of the system obeys Markovian time evolution given by the master equation
\[
         \frac{d}{dt}\rho(t) = \L\rho(t) = -i[\mathcal{H}, \rho(t)] 
         + \sum_{m=1}^{S} \big( 2\mathcal{D}_m\rho(t)  \mathcal{D}^{\dagger}_m 
        - \{\,  \mathcal{D}^{\dagger}_m  \mathcal{D}_m, \rho(t) \, \} \big) \,,
\]
\begin{equation}
        + \sum_{m=S+1}^{S+S'} \big( 2\mathcal{D}^{\dagger}_m\rho(t)  \mathcal{D}_m 
        - \{\,  \mathcal{D}_m  \mathcal{D}^{\dagger}_m, \rho(t) \, \} \big) 
\label{ME1}
\end{equation}
where Hamiltonian $\mathcal{H} \in \D$ is a Hermite operator and $\mathcal{D}_m \in \D$ are arbitrary. \cite{BP, GKS, L, D2}
The (super-)operator $\L$ which acts on $\D$ is called Liouvillian.    

We focus on  Hamiltonian $\mathcal{H}$ of the form
\[
                            \mathcal{H} = \sum_{j, k = 1}^{\n} H_{jk} c^{\dagger}_jc_k 
\]
with Hermite matrix $H$.  
For the operators $\mathcal{D}_m$ which express connection of the system to baths, 
we only consider the form 
\[
    \mathcal{D}_m = \sum_{j=1}^{\n}\overline{\ell_{mj}}c_j  \,. 
    \qquad (\ell_{m1}, \cdots, \ell_{m\n} \in \C)  \,.
\] 
The symbol $\dagger$ is used in two ways: adjoint of the operators acting on $\F$ and Hermite conjugate of matrices.   
Hereafter, we adopt following conventions that
$\C^{\n}$ is the set of all ``column''  complex $\n$-vectors and that  
$( \,\; , \; )$ is the sesqui-linear form on $\C^{\n}$ such that 
\[
           ( \ell_1, \ell_2) = \sum_{j=1}^{\n}\overline{\ell_{1j}}\ell_{2j} = \ell_1^{\dagger}\ell_2
       \qquad \qquad \mbox{for}\quad \ell_i 
       = {}^t(\ell_{i1}, \cdots, \ell_{i\n}) \in \C^{\n},  \quad i =1,2 \,,
\]
and slightly abuse notations like 
\[
          \mathcal{D}_m = \sum_{j=1}^{\n}\overline{\ell_{mj}}c_j  = (\ell_m, c)    \qquad \mbox{and}
     \qquad \mathcal{H} = \sum_{j, k = 1}^{\n} H_{jk} c^{\dagger}_jc_k = (c, Hc) \,.
\]
We will also use the multi-index convention: 
\begin{equation}\label{multiindex}
       (c, \ell)^{\nu} = (c, \ell_1)^{\nu_1} \cdots (c, \ell_{\n})^{\nu_{\n}}  \quad \mbox{ for } 
\quad \nu = (\nu_1, \cdots, \nu_{\n}) \in \{ 0, 1 \}^{\n} \,,
\end{equation}
and so on.  
\begin{defn}\label{def1}
For complex $\n\times\n$ matrix $A$, we define operators $\hL(A), \hG(A), \hF(A)$ and $\hB(A)$ acting on $\D$ by 
\[
       \hL(A)\rho = \sum_{j,k =1}^{\n}A_{jk}c_k\rho c^{\dagger}_j,  \quad 
       \hG(A)\rho = \sum_{j,k =1}^{\n}A_{jk} c^{\dagger}_j\rho c_k, \quad 
       \hF(A)\rho = (c, Ac)\rho
\]
and
\[ \quad \hB(A)\rho = \rho(c, Ac),   \qquad    \mbox{for}  \qquad     \rho\in \D.
\] 
\end{defn} 

\bigskip

The linearity of those operators are obvious. 
They have the following commutation relations. 
\begin{lem}\label{lem1}
For complex $\n\times\n$ matrices $C$ and $D$, the commutation relations  

\begin{align}
     [\hF(C), \hF(D)] = \hF([C,D]), \quad &  [\hB(C), \hB(D)] = - \hB([C,D]), \\
     [\hF(C), \hL(D)] = - \hL(DC), \quad &  [\hB(C), \hL(D)] = - \hL(CD),   \\
      [\hF(C), \hG(D)] = \  \hG(CD), \quad &  [\hB(C), \hG(D)] = \ \hG(DC),
\end{align}
and
\begin{align}
    [\hF(C), \hB(D)] = & [\hL(C), \hL(D)] =  [\hG(C), \hG(D)] = 0,  
\label{FBLG0}\\
    [\hL(C), \hG(D)] & = \tr(CD)  -  \hF(DC) - \hB(CD) 
\end{align}
hold, where ``$\tr$" means the trace for $\n\times\n$ matrix.  
\end{lem}
{\sl Proof} : For $\rho\in\D$, we have
\begin{align*}
      [\hF(C&),  \hF(D)] \rho =  \hF(C) \hF(D)\rho - \hF(D) \hF(C)\rho 
\\
       = &\big((c, Cc)(c, Dc) - (c, Dc)(c, Cc)\big)\rho
      = [(c, Cc), (c, Dc)]\rho 
\\
     & = (c, [C, D]c)\rho  = \hF([C, D])\rho \,,
\end{align*}
where we have used $ [(c, Cc), (c, Dc)] = (c, [C, D]c)$ which is a direct consequence of  (\ref{CAR}).  
Similarly, 
\[
         [\hB(C), \hB(D)] \rho = \rho \big((c, Dc)(c, Cc) - (c, Cc)(c, Dc)\big)  
         = \rho (c, [D, C]c)  = - \hB([C, D])\rho  
\]
is derived.  By using $ [ (c, Cc), c_k] = - (Cc)_k $, we get  
\begin{align*}
      [\hF&(C), \hL(D)]\rho =  (c, Cc)\sum_{j,k}D_{jk}c_k\rho c^{\dagger}_j -
      \sum_{j,k}D_{jk}c_k(c, Cc)\rho c^{\dagger}_j
\\
     &= \sum_{j,k}D_{jk}[ (c, Cc), c_k]\rho c^{\dagger}_j
      = - \sum_{j,k}D_{jk}(Cc)_k\rho c^{\dagger}_j =  - \hL(DC)\rho \,.  
\end{align*}
We also have the other relations:
\begin{align*}
          [\hB&(C), \hL(D)]\rho =  \sum_{j,k}D_{jk}c_k\rho c^{\dagger}_j (c, Cc)
          - \sum_{j,k}D_{jk}c_k\rho(c, Cc) c^{\dagger}_j
\\        
           &=    \sum_{j,k}D_{jk}c_k\rho [c^{\dagger}_j, (c, Cc)] 
          = -\sum_{j,k,l}D_{jk}c_k\rho c^{\dagger}_lC_{lj}  =  - \hL(CD)\rho \,,
\\
           [\hF&(C), \hG(D)] \rho = (c, Cc) \sum_{j,k}D_{jk} c^{\dagger}_j\rho c_k
                     -  \sum_{j,k}D_{jk} c^{\dagger}_j(c, Cc)\rho c_k
\\
             &=     \sum_{j,k}D_{jk} [(c, Cc), c^{\dagger}_j]\rho c_k
             =    \sum_{j,k,l}D_{jk} c^{\dagger}_lC_{lj}\rho c_k      
               =   \hG(CD) \rho \,, 
\\
        [\hB&(C), \hG(D)]  \rho =  \sum_{j,k}D_{jk} c^{\dagger}_j\rho c_k (c, Cc)
        -  \sum_{j,k}D_{jk} c^{\dagger}_j\rho  (c, Cc)c_k
\\      
         &= \sum_{j,k}D_{jk} c^{\dagger}_j\rho [c_k, (c, Cc)]
        =  \sum_{j,k}D_{jk} c^{\dagger}_j\rho (Cc)_k 
        =  \hG(DC) \rho \,.
\end{align*}
The relations (\ref{FBLG0}) are obvious.  
The last relation is derived as
\begin{align*}
            [\hL(C), \hG(D)] \rho =&  \sum_{j,k,l,m}C_{jk} D_{lm}\big(
             c_k c^{\dagger}_l \rho c_mc^{\dagger}_j 
            - c^{\dagger}_l c_k\rho  c^{\dagger}_jc_m \big)
\\
              = \sum_{j,k,l,m}&C_{jk} D_{lm}\big(
             \{c_k, c^{\dagger}_l \}\rho c_mc^{\dagger}_j 
            - c^{\dagger}_l c_k\rho  \{c_m, c^{\dagger}_j\} \big)
\\
         =  \sum_{j,k,l,m}&C_{jk} D_{lm}\big(
             \delta_{kl}\rho (\delta_{mj}-c^{\dagger}_jc_m) 
            - c^{\dagger}_l c_k\rho  \delta_{mj} \big)
\\
                  &= \big(\tr(CD) - \hB(CD) -  \hF(DC)  \big)\rho \,.   \tag*{$\square$}
\end{align*}
Using the notations in Definition \ref{def1}, the Liouvillian in (\ref{ME1}) is expressed in the form
\begin{equation}
      \L = 2\hL(D) + 2\hG(E) + \hF(-iH - D + E) + \hB(iH - D + E) - 2\tr E   \,, 
\label{Liou1}
\end{equation}
where we have put  
\[
       D = \sum_{m=1}^M \ell_m\ell_m^{\dagger}, \qquad 
       E =\sum_{m= M+1}^{M+M'}\ell_m\ell_m^{\dagger}  \,.
\]
Contrary, if $H$ is Hermite and $D$ and $E$ are positive $\n\times\n$ matrices, the operator (\ref{Liou1}) deserves of the Liouvillian of the master equation (\ref{ME1}).  

Let us direct our attention to  the dependence of operator (\ref{Liou1}) on matrices $-iH - D - E$ and $E$ and write it as $\L(-iH -D -E, 2E)$.  
With a slight generalization, we introduce the notation: 
\begin{equation}
    \L(A, M) = \hL( - A - A^{\dagger} - M) + \hG(M) 
    + \hF( A + M) + \hB(A^{\dagger} + M)  -\tr M\,. 
\label{LAM}
\end{equation}
for any complex $\n\times\n$ matrices $A$ and $M$.  
If $A$ and $M$ satisfy $ O \leqslant M \leqslant -A - A^{\dagger} $,  $\L(A, M)$ deserves of the  Liouvillian of (\ref{ME1}).   
In fact, by taking $E = M/2, D = (-A-A^{\dagger} - M)/2$, and $ H =(A^{\dagger} - A)/2i$, we revive (\ref{Liou1}) with $ E, D \geqslant O$ and Hermetian $H$. 

\begin{thm}\label{tLAM} For complex $\n\times\n$ matrices $A, B, M$ and $N$, the following 
commutation relations holds:
\begin{equation}
    [\L(A, M), \L(B, N)] = \L([A, B], AN + NA^{\dagger} - BM- MB^{\dagger}) \,. 
\label{cLAM}
\end{equation}
\end{thm}
{\sl Proof }: Put $ \hS = \hF + \hB - \hL + \hG$. 
Then, from Lemma \ref{lem1}, we get the following commutation relations by straight forward calculations: 
\begin{align*}
&[(\hF-\hL)(C), (\hF- \hL)(D) ] = (\hF- \hL)([C,D])\,,  \\
&[(\hB - \hL )(C), (\hB - \hL)(D) ] = - (\hB- \hL)([C,D])\,, \\
&[(\hF - \hL )(C), (\hB - \hL)(D) ] = 0\,, \\
&[(\hF - \hL )(C), \hS(D) ] = \hS(CD) - \tr(CD)\,,  \\
&[(\hB - \hL )(C),  \hS(D) ] = \hS(DC) - \tr(DC)\,, \\ 
&[ \hS(C),  \hS(D)] =0 \,.  
\end{align*}
Then we get
\begin{align*}
        &  [\L(A,M), \L(B, N) ]    \\
        = [ (\hF-\hL&)(A)  + (\hB - \hL )(A^{\dagger}) + \hS(M) \,, 
             (\hF-\hL)(B) + (\hB - \hL )(B^{\dagger}) + \hS(N) ] \\
             =  (\hF- \hL)&([A, B]) + \hS(AN) - \tr(AN) - (\hB- \hL)([A^{\dagger},B^{\dagger}]) \\
             &+\hS(NA^{\dagger}) - \tr (NA^{\dagger})
                -\hS(BM) + \tr (BM) -\hS(MB^{\dagger})  + \tr(MB^{\dagger})\\
             = (\hF-\hL)&([A, B]) + (\hB - \hL )([A, B]^{\dagger}) 
             + \hS(AN+NA^{\dagger}-BM -MB^{\dagger})   \\
             & - \tr(AN + NA^{\dagger} - BM -MB^{\dagger}) \\
         =  \L([A, B]&, AN + NA^{\dagger} - BM- MB^{\dagger}) \,.  \tag*{$\square$}
\end{align*}

Since $\F$ is a finite dimensional Hilbert space, 
there are nothing delicate or difficult in the analysis 
of $\D$ (i.e., the space of operators acting on $\F$)  
and of the operators acting on $\D$.       
All operators we consider in this paper are everywhere defined 
bounded operators.  
Each of them is able to generate a one parameter (semi-)group.  
Due to \cite{GKS, L, D}, $\L( A, B) $ is a generator of completely positive 
and trace preserving semigroup, if 
$O \leqslant  M \leqslant -A -A^{\dagger}$.  
For general $ A, M $, $\L(A, M)$ generates a trace preserving semigroup.  
Indeed, we can derive $\Tr \L(A, M)\rho = 0$ for eny $\rho \in \D$, readily.  
Here $\Tr$ is the trace operation for operators on $\F$.        

Note also that $\L( A, M)$ is linear and continuous in terms of 
$( A, M) \in \M(\n, \C)^2$, i.e., 
\[ \alpha_1\L( A_1, M_1) + \alpha_2\L( A_2, M_2) 
      =  \L( \alpha_1A_1 + \alpha_2A_2, \, \alpha_1M_1 + \alpha_2M_2) \,,
\]
\[
    \lim_{n\to\infty}\L( A_n, M_n ) = 
    \L( \lim_{n\to\infty}A_n, \lim_{n\to\infty}M_n) \,.  
\]
Indeed, $\L(A, M)$ is a linear combination of bounded operators on 
$\D$ and its coefficients are components of the matrices $A$ and $M$.   
It would not be needed to give detailed explanations on these 
analysis on finite dimensional spaces, here and hereafter.   
For these subjects, we refer to basic textbooks such as \cite{Di}.

\section{Affine transformations}\label{AffineT}
In this section, we see that the commutation relation in Theorem \ref{tLAM} is related to certain affine transformation.  
Hereafter we use the following notations: 
\begin{align*}
                \M(\n;\C) &: \mbox{the vector space consists of all complex $\n\times\n$ matrices;}
\\
               GL(\n;\C) &: \mbox{the general linear group consist of all invertible elements in }
               \M(\n;\C) \,.
\end{align*}
We consider the following affine transformation on $\M(\n;\C)$ 
\begin{equation}\label{AT}
             (U, M) : X \mapsto UXU^{\dagger} + M 
\end{equation}
induced by $U \in GL(\n;\C)$ and  $ M \in  \M(\n;\C)$.  
The set of these transformations may be regarded as an action 
of the semidirect product group
$ \G = GL(\n;\C) \ltimes  \M(\n;\C) $ whose group law is given by 
\begin{equation}
                    (U, M) \circ (V, N) = (UV, UNU^{\dagger} + M) \,.
\label{grouplaw}
\end{equation}

Now, we consider the Lie algebra of the semidirect product.   
For the task, we must consider the space of smooth functions on $\G$, 
$C^{\infty}(\G)$, and the representation of the group on it: 
\begin{equation}
              \big(r(U,M)f\big)(V, N) = f\big( (U,M)^{-1}(V, N) \big) = 
               f\big(U^{-1}V, U^{-1}(N-M)U^{\dagger -1} \big)
\label{r}
\end{equation}
for $f \in C^{\infty}(\G)$ and $(U, M), (V, N) \in \G$.  
Then, the Lie algebra of $ \G $ is the set of all differential operators on $ C^{\infty}(\G)$ of the form 
\begin{equation}
        (\!( A, M)\!) = \lim_{\epsilon \to 0}\frac{r(I+\epsilon A, \epsilon M) - r(I, O)}    
        {\epsilon} \qquad  (A,M \in \M(\n;\C) ) \,.
\label{rgenerator}
\end{equation}
Since $r$ is homomorphism,   
\begin{align*}
      &\big( r(I+\epsilon A, \epsilon M) - r(I, O)\big)
      \big(r(I+\epsilon B, \epsilon N) - r(I, O)\big)f  \\
      & =  \big(r(I+\epsilon( A + B) + \epsilon^2 AB, \epsilon( M + N) 
      + \epsilon^2(AN + NA^{\dagger}))     \\
      &-  r(I+\epsilon A, \epsilon M)
      -  r(I+\epsilon B, \epsilon N) + r(I, O)\big)f  + O(\epsilon^3)
\end{align*}
therefore
\begin{align*}
      &[ r(I+\epsilon A, \epsilon M) - r(I, O),
      r(I+\epsilon B, \epsilon N) - r(I, O)]f  \\
      & =  \Big(r \big(I+\epsilon( A + B) + \epsilon^2 AB, \epsilon( M + N) 
      + \epsilon^2(AN + NA^{\dagger}) \big)     \\
      &-  r \big(I+\epsilon (A + B) + \epsilon^2 BA, \epsilon( M + N) 
      + \epsilon^2(BM + MB^{\dagger}) \big)\Big)f  + O(\epsilon^3)\\
      &=  \Big(r  \big(I + \epsilon^2 [A, B],  
      \epsilon^2(AN + NA^{\dagger} -BM - MB^{\dagger})  \big) - r(I, O) \Big)     \\
      &\Big( r \big(I+\epsilon (A + B) + \epsilon^2 BA, \epsilon( M + N) 
      + \epsilon^2(BM + MB^{\dagger})  \big)\Big)f  + O(\epsilon^3)
\end{align*}
hold.   
Dividing by  $\epsilon^2$ and taking the limit $\epsilon \to 0$, we obtain the commutation relation
\begin{equation}
          [ (\!(A, M)\!), (\!(B,N)\!)] = (\!([A, B], AN + NA^{\dagger} - BM - MB^{\dagger} )\!)  \,.  
\label{rcomm}
\end{equation}
Comparing with (\ref{cLAM}), we understand that   
the algebra of Liouvillians $\{ \, \L(A, M) \, | \, A, M \in \M(\n;\C) \, \}$ and 
the Lie algebra of the affine transformation group $ \G $ are
isomorphic.   

Because isomorphism between Lie algebras of two Lie groups does not always imply 
the isomorphism between the Lie groups themselves (e.g., $SU(2)$ and $SO(3)$),  
a relation mong elements in $ \G $ does not ensure the corresponding relation among  evolution operators on density matrices.    
However, the former may gives motivation and vision to the study of the latter.   

Consider
\[
              p_t(A, M) = \Big(e^{tA}, \int_0^t e^{sA}Me^{sA^{\dagger}}\, ds \Big) 
\]
for $A, M \in \M(\n;\C)$ and $t\geqslant 0$. 
It is derived easily from (\ref{grouplaw}) that
\[
       p_t(A, M) \circ p_s(A, M) = p_{t+s}(A, M)   \qquad \mbox{for}  \quad t,s \geqslant 0 \,.  
\]
It is also obvious that $\lim_{t\downarrow 0}p_t(A, M) = p_0(A, M) = (I, O)$ and 
\[
           \Big(\frac{d}{dt}\Big)_{t=0}rp_t(A, M) = (\!( A, M)\!) \,.  
\] 
Thus, $\{ rp_t(A, M) \}_{t\geqslant 0}$ is a one parameter semigroup with generator  $(\!( A, M)\!)$.   
In this sense, the notation $ e^{t (\!( A, M)\!)}  = rp_t(A,M) $ is plausible.  
However we may denote
\[
                  e^{t (\!( A, M)\!)}  = p_t(A,M)
\]
without confusion.   
\begin{lem} The following formulae hold for $A, B, M, N \in \M(\n;\C)$: 
  \begin{align*}
  &(1) \hskip1cm  e^{t (\!( A, M)\!)} = \Big(e^{tA}, \int_0^t e^{sA}Me^{sA^{\dagger}}\, ds \Big) \,;\\
  &(2) \hskip1cm  e^{ (\!( O, M)\!)} = (I, M) \,, \quad  
         e^{ (\!( A, O)\!)} = (e^A, O) \ ; \\ 
  &(3) \hskip1cm    e^{t (\!( A, M)\!)} = e^{ (\!( O, \, \int_0^t e^{sA}Me^{sA^{\dagger}} ds)\!)} \circ e^{t (\!( A, O)\!)} \,; \\
  &(4)  \hskip1cm    e^{t (\!( A, M)\!)} =  e^{(\!( O, T )\!)} \circ e^{t (\!( A, O)\!)} \circ e^{ - (\!( O, T)\!)}   \quad \mbox{if}\quad  AT +TA^{\dagger} = - M  \,.
  \end{align*}
\label{formulae1}
\end{lem}
{\sl Proof :} (1) has been shown above.  
(2) exhibits special cases of (1).  

\noindent(3) Because of (1) and (2), (3) is derived from (\ref{grouplaw}).  

\noindent(4) Note that 
\[
              e^{tA}Te^{tA^{\dagger}} - T = \int_0^t \frac{d}{ds} e^{sA}Te^{sA^{\dagger}} \,ds
              = \int_0^t e^{sA}(AT + TA^{\dagger})e^{sA^{\dagger}} \,ds = -\int_0^t e^{sA}Me^{sA^{\dagger}} \,ds  \,. 
\]
We get
\[
    \mbox{r.h.s.\,of (4)} = (I, T)\circ(e^{tA}, O) \circ(I, -T) 
    = (e^{tA},  - e^{tA}Te^{tA^{\dagger}} + T)
\]
\[
    = \big( e^{tA}, \int_0^t e^{sA}Me^{sA^{\dagger}} \,ds \big)= \mbox{l.h.s.of (4)}   \,. \hskip3cm  \square
\]
\begin{rem}\label{restriction}
Let $ \H(\n;\C) $ be the subspace of $ M(\n;\C)$ consists of all Hermite elements therein.  
If we choose $ M \in \H(\n;\C)$, (\ref{AT}) maps $\H(\n;\C)$ into $\H(\n;\C)$.  
Correspondingly, $ GL(\n;\C) \ltimes \H(\n;\C) $ is a subgroup of
$ \G = GL(\n;\C) \ltimes \M(\n;\C) $.  
Hence by restriction, we may confer Lemma\ref{formulae1} as a set of formulae concerning $ GL(\n;\C) \ltimes \H(\n;\C) $ and its Lie algebra.  
\end{rem}

\section{Time evolution}\label{timeevol}
From the isomorphism between the Lie algebra of Liouvillians (\ref{LAM}) and that associated with the affine transformation group $\G$, it is expected that time evolutions induced by those Liouvillians have the properties corresponding to Lemma \ref{formulae1}.  
In this section, however, we prove the properties of time evolution directly from the commutation relation (\ref{cLAM}), without consulting expected connection to Lemma \ref{formulae1}.   
\begin{thm}\label{thmTimeEvol} For $A, M \in \M(\n;\C) $, 
\[
           e^{t\L(A,M)} = e^{\L( O, \int_0^te^{sA}Me^{sA^{\dagger}}ds)}
      e^{t\L(A,O)}
\]
holds.  
\end{thm}
\begin{lem}
\label{com2}
For $A, M, T \in \M(\n;\C) $, the following formulae hold. 
\begin{description}
   \item[ \rm(1)] \qquad 
$\displaystyle  e^{t\L( A, O)} \L( O, M) 
             = \L( O, e^{tA}Me^{tA^{\dagger}})e^{t\L( A, O)} $
\item{\rm(2)} \qquad  $\displaystyle e^{\L( O, T)} \L( A, M) e^{-\L( O, T)} 
             = \L( A, M -AT -TA^{\dagger}) $
\end{description}
\end{lem}
{\sl Proof : } (1) Put 
\[
    f(s) =  e^{(t-s)\L( A, O)}  \L( O, e^{sA}Me^{sA^{\dagger}})e^{s\L( A, O)}.   
\]
Then from (\ref{cLAM}), we have 
\[
          f'(s) =  e^{(t-s)\L( A, O)} \Big( - [\L( A, O),  \L( O, e^{sA}Me^{sA^{\dagger}})] 
          + \L( O, e^{sA}(AM + MA^{\dagger})e^{sA^{\dagger}}) \Big)e^{s\L( A, O)} = 0\,,
\]
which implies $f(0) = f(t)$.  Thus the desired equality is derived.   

(2)   Put   
\[
    g(s) =  e^{s\L( O, T)}  \L( A, M)e^{-s\L( O, T)}.   
\]
Then we have $g(0) = \L( A, M)$ and 
\[
        g'(s) =  e^{s\L( O, T)} [ \L( O, T), \L( A, M)]e^{-s\L( O, T)}
\]
\[
                = e^{s\L( O, T)}  \L( O, -AT- TA^{\dagger}) e^{-s\L( O, T)}
                = \L( O, -AT- TA^{\dagger}).   
\]
Thus we get $g(1) = \L( A, M -AT- TA^{\dagger})\,$.  
Here we note $ [\L( O ,M) , \L( O, N) ] =0 $\,. \hfill$\square$

{\bf Proof of Theorem \ref{thmTimeEvol}} \quad
Put  
\[
              h(s) =  e^{\L( O, \int_0^se^{uA}Me^{uA^{\dagger}}du)}
      e^{s\L(A,O)}e^{(t-s)\L(A,M)} \,.
\]
We have
\[
           h'(s) =  e^{\L( O, \int_0^se^{uA}Me^{uA^{\dagger}}du)}
      \Big( \L( O, e^{sA}Me^{sA^{\dagger}} ) e^{s\L(A,O)}  -  e^{s\L(A,O)}\L( O, M)
      \Big)e^{(t-s)\L(A,M)} 
\]
\[
          =   e^{\L( O, \int_0^se^{uA}Me^{uA^{\dagger}}du)}
      \Big( \L( O, e^{sA}Me^{sA^{\dagger}} ) e^{s\L(A,O)} -  \L( O, e^{sA}Me^{sA^{\dagger}})
       e^{s\L(A,O)} \Big)e^{(t-s)\L(A,M)} = 0 \,,  
\]
where we have used the linearity $\L(A, O) -\L(A, M) = - \L( O, M)$ and 
Lemma \ref{com2}(1).    
Therefore we get $h(0) = h(t)$ which is the desired formula.   \hfill $\square$
\subsection{Basis of $\D$}\label{sectionBbase}
\begin{defn}\label{defnBase}
For $\xi_1, \cdots, \xi_{n}, \eta_1, \cdots, \eta_{m} \in \C^{\n}$, 
we define the following elements of $\D$:
\begin{equation}
     \Phi(\xi_1, \cdots, \xi_n; \eta_1, \cdots, \eta_m) = 
  \begin{cases}
      \L( O, \xi_1\eta_1^{\dagger})\cdots\L( O, \xi_n\eta_n^{\dagger})
        \big(\Omega(\eta_m, c)\cdots(\eta_{n+1}, c)\big)  \quad (n \leqslant m)   \\
      \L( O, \xi_1\eta_1^{\dagger})\cdots\L( O, \xi_m\eta_m^{\dagger})
       \big((c, \xi_{m+1}) \cdots (c, \xi_n) \Omega\big)  \quad ( n > m)
  \end{cases}, 
\label{Base}
\end{equation}
and
\begin{equation}
     \Pi(\xi_1, \cdots, \xi_n; \eta_1, \cdots, \eta_m) = 
        (c, \xi_{1}) \cdots (c, \xi_n) \Omega(\eta_m, c)\cdots(\eta_{1}, c)   \,,
\label{sBase}
\end{equation}
where $ n, m = 0,1,2, \cdots, \n$.   
\end{defn}

\begin{rem}\label{inducphi}
     For $n=0$ or $m=0$, (\ref{Base}) and (\ref{sBase}) mean that
\begin{align*}
          \Phi( \emptyset; \emptyset ) &= \Pi( \emptyset; \emptyset ) 
          \ = \ \Omega \,,
\\
             \Phi(\emptyset ; \eta_1, \cdots, \eta_m) &= 
             \Pi(\emptyset ; \eta_1, \cdots, \eta_m)  
              =   \Omega(\eta_m, c)\cdots(\eta_{1}, c)   \,, 
\\
             \Phi(\xi_1, \cdots, \xi_n; \emptyset ) &=  
             \Pi(\xi_1, \cdots, \xi_n; \emptyset ) 
             \ =   (c, \xi_{1}) \cdots (c, \xi_n) \Omega  \,.  
\end{align*}
Note that for $ n, m \geqslant 1$, 
\begin{equation}\label{r44}
         \Phi(\xi_1, \xi_2, \cdots, \xi_n; \eta_1, \eta_2, \cdots, \eta_m) =  
         \L( O, \xi_1\eta_1^{\dagger}) \Phi(\xi_2, \cdots, \xi_n; \eta_2, \cdots, \eta_m)  
\end{equation}
holds.  
\end{rem}

\begin{prop}\label{basisprop}
\begin{description}
   \item[\rm(1)]\ The element $\displaystyle \Omega(\xi_1, \cdots, \xi_n; \eta_1, \cdots, \eta_m)$ is anti-symmetric under permutations of $\xi_1, \cdots, \xi_n $ and of $\eta_1, \cdots, \eta_m $, i.e., 
\begin{equation}
  \Phi(\xi_{\sigma(1)}, \cdots, \xi_{\sigma(n)}; \eta_{\tau(1)}, \cdots, \eta_{\tau(m)}) 
  = \sgn(\sigma)\sgn(\tau)\Phi(\xi_1, \cdots, \xi_n; \eta_1, \cdots, \eta_m)
\label{perm}
\end{equation}
holds.      
  \item[\rm(2)] \ The equalities
\[
   \Phi(\xi_1, \cdots, \xi_n; \eta_1, \cdots, \eta_m) = 
   \sum_{ \sigma \in \mathfrak{S}_n, \tau \in \mathfrak{S}_m}\sum_{p=0}^{\min\{n,m\}}
    (-1)^{p} \sgn(\sigma)\sgn(\tau)
\]
\begin{equation}
    \times \ \frac{ \prod_{j=1}^p\big(\eta_{\tau(j)}, \xi_{\sigma(j)}\big)}{p!}          \frac{\Pi(\xi_{\sigma(p+1)}, \cdots, \xi_{\sigma(n)} ; 
 \eta_{\tau(p+1)}, \cdots, \eta_{\tau(m)} )}{(n-p)! \ (m-p)!} 
\label{OmegaPi}
\end{equation}
and
\[
     \Pi(\xi_1, \cdots, \xi_n; \eta_1, \cdots, \eta_m) = 
   \sum_{ \sigma \in \mathfrak{S}_n, \tau \in \mathfrak{S}_m}\sum_{p=0}^{\min\{n,m\}}
     \sgn(\sigma)\sgn(\tau)
\]
\begin{equation}
    \times \ \frac{ \prod_{j=1}^p\big(\eta_{\tau(j)}, \xi_{\sigma(j)}\big)}{p!}          \frac{\Phi(\xi_{\sigma(p+1)}, \cdots, \xi_{\sigma(n)} ; 
 \eta_{\tau(p+1)}, \cdots, \eta_{\tau(m)} )}{(n-p)! \ (m-p)!} 
\label{PiOmega}
\end{equation}
hold.  
  \item[\rm(3)] \ If both of $\{\xi_i \, | \,i=1,\cdots, \n \, \}$ and $\{\, \eta_j \, | \,j=1,\cdots, \n \, \} $ are bases of $\C^{\n}$, then 
\[
  \Big\{\Phi(\xi_{i_1}, \cdots, \xi_{i_n}; \eta_{j_1}, \cdots, \eta_{j_m}) \, \Big| \, 
     0 < i_1 < i_2 < \cdots <i_n \leqslant \n, 
\]
\begin{equation}
     \ 0 < j_1 < j_2 < \cdots < j_m \leqslant \n \,,    \quad  0 \leqslant n, m \leqslant \n \,   \Big\}
\label{omegabase}
\end{equation}
is a basis of $\D$.
\end{description}
\end{prop}
{\sl Proof : } For (1), we first show the following properties: 
\begin{description} 
\item[\rm{(i)}] $ \{(c, \xi_1), (c, \xi_2) \} = \{ (\eta_1, c), (\eta_2, c) \}=  0 $ ;
\item[\rm{(ii)}] $ [ \L(O, \xi_1\eta_1^{\dagger}), \L(O, \xi_2\eta_2^{\dagger})] = 0 $ ;
\item[\rm{(iii)}] $ \L( O, \xi_1\eta^{\dagger} )\big((c,\xi_2) \Phi(\xi_3, \cdots, \xi_n; \emptyset ) \big)  + \L( O, \xi_2\eta^{\dagger} ) \big((c, \xi_1) \Phi(\xi_3, \cdots, \xi_n; \emptyset ) \big)  = 0 $ ;
\item[\rm{(iv)}] $ \L( O, \xi\eta_1^{\dagger})\big( \Phi(\emptyset ; \eta_3, \cdots, \eta_m)(\eta_2,c) \big) + \L( O,  \xi\eta_2^{\dagger}) \big( \Phi(\emptyset ; \eta_3, \cdots, \eta_m) (\eta_1,c) \big) = 0 $ ;
\item[{(v)}] $ \L( O ,\xi_1\eta_1^{\dagger})\L( O ,\xi_2\eta_2^{\dagger}) + \L( O ,\xi_2\eta_1^{\dagger})\L( O ,\xi_1\eta_2^{\dagger})= 0 $\,.
\end{description}
The first two properties are direct consequences of (\ref{CAR}) and (\ref{cLAM}), respectively.  
For the other ones, we note that 
\begin{align}
               \L( O, \xi\eta^{\dagger} ) \rho = & -  (\eta, c) \rho (c, \xi) 
               +  (c, \xi)  \rho (\eta, c) +  \{ (c, \xi) (\eta, c), \rho \}    
                - (\eta, \xi) (\eta, c)\rho  \notag \\
                & =  [ (c, \xi) , \{ (\eta, c) ,  \rho \} ] 
                = [ \{ (c, \xi) , \rho \},  (\eta, c) ] \,.
\label{LOM}
\end{align}
Then, (iii) is derived as 
\[
     \L( O, \xi_1\eta^{\dagger} ) \big( (c,\xi_2) \Phi(\xi_3, \cdots, \xi_n; \emptyset ) \big)
      = [ \{ (c, \xi_1) ,  (c,\xi_2) \Phi(\xi_3, \cdots, \xi_n; \emptyset ) \},  (\eta, c) ]
\]
\[     = [  (c, \xi_1)  (c,\xi_2) \Phi(\xi_3, \cdots, \xi_n; \emptyset ) ,  (\eta, c) ] 
      = - [  (c, \xi_2)  (c,\xi_1) \Phi(\xi_3, \cdots, \xi_n; \emptyset ) ,  (\eta, c) ] 
\]
\[
    = - \big(\L( O, \xi_2\eta^{\dagger} )(c,\xi_1) \Phi(\xi_3, \cdots, \xi_n; \emptyset ) \,.
\]
Similarly, we have (iv).  For (v), we begin from
\begin{equation}
   \L( O, \xi\eta^{\dagger} )^2 = 0 . 
\label{L20}
\end{equation}
It is derived as 
\[
       \L( O, \xi\eta^{\dagger} )^2 \rho  
       =  [ (c, \xi) , \{ (\eta, c) , \L( O, \xi\eta^{\dagger} ) \rho \}]
       =   [ (c, \xi) , \{ (\eta, c) ,  [ \{ (c, \xi) , \rho \},  (\eta, c) ] \} ] 
\]
\[
        =   \Big[ (c, \xi) , \Big( (\eta, c) \{ (c, \xi) , \rho \} (\eta, c)  
        -  (\eta, c) [ \{ (c, \xi) , \rho \} (\eta, c) \Big) \Big] = 0 \,, 
\]
since $(\eta, c)^2 = 0$. 
By putting $ \xi = \xi_1 + \xi_2$ in (\ref{L20}) and using (ii) and (\ref{L20}) itself, we have 
$ \L( O, \xi_1\eta^{\dagger} )\L( O, \xi_2\eta^{\dagger} ) =0 $\,.  
By putting $ \eta = \eta_1 +  \eta_2 $ in this equality, we get (v) similarly.  

\medskip

It is enough to prove (1) for the cases that $\sigma$ is any transposition of neighboring two elements and  $\tau$ is the identity and vice versa.  
Let us consider the case $ n>m$ and $\sigma = ( j, j+1)$ and $\tau$ is the identity.  
If $ m < j < n$,  (\ref{perm}) is a consequence of (i).  
If $ 2 \leqslant j +1  \leqslant m  $, (v) implies  (\ref{perm}).    
If $ j =m$, (iii) implies   (\ref{perm}).  
The other cases are shown similarly.  

(2) Due to multi-linearity of $\Phi(\xi_1, \cdots, \xi_n; \eta_1, \cdots, \eta_m)$ with respect to $ \xi_1, \cdots, \xi_n$ and $\eta_1^{\dagger}, \cdots, \eta_m^{\dagger} $, 
it is enough to prove (\ref{OmegaPi}) for the case where $\xi_i$'s and $\eta_j$'s are elements of the standard basis $\{ e_j \}$ of $\C^{\n}$.  
Moreover, due to the anti-symmetricity (1), it is enough to consider the case 
without loss of generality
\[
     \Phi(e_1, \cdots, e_q,e_{q+1}, \cdots, e_n; e_1, \cdots, e_q,e_{n+1}, \cdots, e_{n+m-q}) = 
   \sum_{ \sigma \in \mathfrak{S}_n, \tau \in \mathfrak{S}_m}\sum_{p=0}^{q}
    (-1)^{p} \sgn\sigma \, \sgn\tau
\]
\begin{equation}\label{PPee}
    \times \ \frac{ \big( \prod_{j=1}^p  \delta_{\tau(j),\sigma(j)}\big)}{p!}          \frac{\Pi(e_{\sigma(p+1)}, \cdots, e_{\sigma(n)} ; 
 e_{\tau(p+1)}, \cdots, e_{\tau(q)}, e_{\tau(n+1)} \cdots e_{\tau(n+m-q)} )}{(n-p)! \ (m-p)!} \,,
\end{equation}
where $\tau$ runs over all permutations of $ \{ 1, \cdots, q, n+1, \cdots, n+m-q \}$.  
The right hand side of (\ref{PPee}) equals to 
\begin{equation}\label{OEE}
      \sum_{p=0}^q (-1)^p \sum_{X \subset \{1, \cdots, q\}, |X| = q-p}
     \Pi\big(e_{x_1}, \cdots, e_{x_{q-p}},e_{p+1}, \cdots, e_n; 
         e_{x_1}, \cdots, e_{x_{q-p}},e_{n+1}, \cdots, e_{n+m - q}\big)
\end{equation}
where we denote $X=\{x_1, \cdots, x_{q-p}\}$.  
Note that the equalities  
\begin{align}
  \L( O, e_{q+1}e_{n+1}^{\dagger}) \Pi &\big(e_{p+2}, \cdots, e_n; 
         e_{n+2}, \cdots, e_{n+m - q}\big)
\notag \\
     & = \Pi\big(e_{q+1}, e_{q+2}, \cdots, e_n; 
         e_{n+1}, e_{n+2}, \cdots, e_{n+m - q}\big) \,, 
\label{a} \\
   \L( O, e_{p}e_{p}^{\dagger}) \Pi\big(e_{p+1}&, \cdots, e_n; 
         e_{n+1}, \cdots, e_{n+m - q}\big)
\notag \\
     = \Pi\big(e_{p}, e_{p+1},& \cdots, e_n; 
         e_p, e_{n+1}, e_{n+2}, \cdots, e_{n+m - q}\big) 
       -  \Pi\big(e_{p+1}, \cdots, e_n; 
         e_{n+1}, \cdots, e_{n+m - q}\big)\,,
\label{b}
\end{align}
and so on hold by (\ref{LOM}) and orthonormality of $e_j$'s.    
It follows from (\ref{r44}) and (\ref{a}) that
\[
        \Phi\big(e_{q+1}, e_{q+2}, \cdots, e_n; 
         e_{n+1}, e_{n+2}, \cdots, e_{n+m - q}\big) 
\]
\begin{equation}\label{c}
         = \Pi\big(e_{q+1}, e_{q+2}, \cdots, e_n; 
         e_{n+1}, e_{n+2}, \cdots, e_{n+m - q}\big)  
\end{equation}
holds.  
And due to (\ref{r44}), (\ref{b}) and (\ref{c}), it is not difficult to see 
that (\ref{OEE}) is equal to the left-hand side of (\ref{PPee}).  

To show (\ref{PiOmega}), we use the expression (\ref{OmegaPi}) of $\Phi$  in the right  hand side of (\ref{PiOmega}), inclusion exclusion principle and anti-linearity of $\Pi$.   
Then the left-hand side derived.   

The claim (3) is a consequence of above (2) and the obvious fact that
\[
  \Big\{\Pi(\xi_{i_1}, \cdots, \xi_{i_n}; \eta_{j_1}, \cdots, \eta_{j_m}) \, \Big| \, 
     0 < i_1 < i_2 < \cdots <i_n \leqslant \n, 
\]
\[
     \ 0 < j_1 < j_2 < \cdots < j_m \leqslant \n \,,    \quad  0 \leqslant n, m \leqslant \n \,   \Big\}
\]
is a basis of $\D$.  
\hfill $\square$

\begin{prop}\label{actA}
For any $\xi_1, \cdots, \xi_{\n}, \eta_1, \cdots, \eta_{\n} \in \C^{\n}$, 
the equalities
\[
 \L(A, O)\Phi(\xi_1, \cdots, \xi_n; \eta_1, \cdots, \eta_m) = \sum_{i=1}^n\Phi(\xi_1, \cdots, A\xi_i, \cdots \xi_n; \eta_1, \cdots, \eta_m)
\]
\begin{equation}\label{dA}
  + \sum_{j=1}^m\Phi(\xi_1, \cdots, \xi_n; \eta_1, \cdots, A\eta_j, \cdots \eta_m)
\end{equation}
and 
\begin{equation}\label{GammaA}
 e^{t\L(A, O)}\Phi(\xi_1, \cdots, \xi_n; \eta_1, \cdots, \eta_m) = \Phi(e^{tA}\xi_1, \cdots, e^{tA}\xi_n; e^{tA}\eta_1, \cdots, e^{tA}\eta_m)
\end{equation}
hold. 
\end{prop}
{\sl Proof : } Let us recall
\begin{equation}
   \L( A, O) \rho = -\sum_{jk}(A_{jk} +A_{jk}^{\dagger})c_k\rho c_j^{\dagger} 
    +(c, Ac)\rho + \rho(c,A^{\dagger}c)  
\label{LAO}
\end{equation}
and commutation relations on $\F$
\[
            [(c, Ac), (c, \xi)] = (c, A\xi), \quad 
       [(\eta, c), (c, A^{\dagger}c)] = (A\eta, c)  \,.
\]
Then we have $ \L( A, O) \Omega = 0$,   
\[
          \L( A, O)\big((c,\xi_1) \cdots (c,\xi_n)\Omega\big) 
          = (c, Ac)(c,\xi_1) \cdots (c,\xi_n)\Omega 
\]
\[
          =  [(c, Ac), (c,\xi_1) \cdots (c,\xi_n)]\Omega 
         = \sum_{\ell} (c,\xi_1) \cdots[(c, Ac), (c,\xi_{\ell})] \cdots (c,\xi_n)\Omega 
\]
\[
        = \sum_{\ell} (c,\xi_1) \cdots(c, A\xi_{\ell})] \cdots (c,\xi_n)\Omega   
\]
and 
\[
         \L( A, O)\big(\Omega(\eta_m,c) \cdots(\eta_1, c) \big)
         = \sum_{\ell} \Omega(\eta_m,c) \cdots(A\eta_{\ell}, c) \cdots(\eta_1, c) \,.   
\]
Combined with the commutation relation 
\[
             [\L( A, O), \L( O, \xi\eta^{\dagger})] 
     = \L( O,  A\xi\eta^{\dagger} + \xi(A\eta)^{\dagger}),  
\]
and (\ref{Base}), we get (\ref{dA}).   
For (\ref{GammaA}), let us denote its right-hand side by $X_t$.  
Then it is obvious to see
\[
       X_0 = \Phi(\xi_1, \cdots, \xi_n; \eta_1, \cdots, \eta_m)\,,   \qquad         \frac{dX_t}{dt} = \L( A, O) X_t \,,
\]
which proves (\ref{GammaA}).   \hfill$\square$

\subsection{Properties of $A$}\label{sectionPropA}
In this subsection, we deal with spectral properties of $A \in \M(\n;\C)$ satisfying $ -A - A^{\dagger} \geqslant O$.   
Let $\lambda_1, \cdots, \lambda_r$ be the distinct eigenvalues of $A$.  
We consider the decomposition
\[
            \C^{\n} = \bigoplus_{j=1}^{r} V_{\lambda_j} \,, 
\]
where $V_{\lambda_j} = \mathrm{Ker\,} (A -\lambda_j)^{\n}$ is the generalized eigenspace of $\lambda_j$. \cite{C}     
The following properties hold.   
\begin{prop}
\begin{description}
   \item[\rm{(1)}]\qquad $\Re \lambda_j \leqslant 0$.  
   
   \item[\rm{(2)}]\qquad If $\Re \lambda_j = 0$, $V_{\lambda_j} = \mathrm{Ker\,}(A-\lambda_j)$ holds. \quad (i.e., $V_{\lambda_j}$ is the eigenspace of $\lambda_j$). 

  \item[\rm{(3)}]\qquad If $\Re \lambda_j = 0$ and $\lambda_k \ne \lambda_j$, 
  $V_{\lambda_j} \perp V_{\lambda_k}$ holds.   
\end{description}
\end{prop}
{\sl Proof : } (1) Due to the assumption $ -A - A^{\dagger} \geqslant O$, we have
\[
  0 \geqslant (\phi, (A+A^{\dagger})\phi) = (\phi, A\phi) +  (A\phi,\phi)= 2\Re \lambda_j \|\phi\|^2 
\]
for $\phi \in V_{\lambda_j}$.   
Thus we have $\Re \lambda_j \leqslant 0$.

(2) Suppose that $V_{\lambda_j} \supsetneqq \mathrm{Ker\,} (A - \lambda_j)$.  
Then, there exist $\phi, \chi \in V_{\lambda_j} - \{ 0 \}$ satisfying $A\phi =\lambda_j\phi$ and $A\chi =\lambda_j\chi + \phi$.   
From $ -A - A^{\dagger} \geqslant O$ and $\Re \lambda_j =0$, 
\[
         0 \geqslant (\phi +\epsilon\chi, (A+A^{\dagger})(\phi +\epsilon\chi)) 
         = 2\epsilon\|\phi\|^2 + 2\epsilon^2\Re(\phi, \chi)
\]
holds for $\epsilon > 0$.  
Dividing it by $\epsilon$ and taking the limit $\epsilon\downarrow 0$, we get a contradiction $\|\phi\|^2 \leqslant 0$.    

(3) For sequence $ \chi_0, \chi_1, \cdots$ in $V_{\lambda_k} - \{ 0 \}$ satisfying $A\chi_0 = \lambda_k\chi_0,  A\chi_1 = \lambda_k\chi_1 + \chi_0, \cdots$.  
Then for any $\phi \in V_{\lambda_j} $, 
\[
      0 \geqslant (\phi +\epsilon e^{i\theta}\chi_0, (A+A^{\dagger})(\phi +\epsilon e^{i\theta}\chi_0)) 
         = 2\epsilon \Re\big( (\lambda_k -\lambda_j)e^{i\theta}(\phi, \chi_0)\big ) + 2 \epsilon^2\Re \lambda_k \|\chi_0\|^2      
\]
holds for $\epsilon > 0$ and $\theta \in \R$.  
Dividing it by $\epsilon$ and taking the limit $\epsilon\downarrow 0$, we get 
\[
  \Re\big( (\lambda_k -\lambda_j)e^{i\theta}(\phi, \chi_0)\big) \leqslant 0\,.
\]
Since $\theta$ is arbitrary, $(\phi, \chi_0) = 0 $ follows.  
Similarly, from 
\[
    0 \geqslant (\phi +\epsilon e^{i\theta}\chi_1, (A+A^{\dagger})(\phi +\epsilon e^{i\theta}\chi_1)) 
         = 2\epsilon \Re\big( e^{i\theta}(\lambda_k -\lambda_j)(\phi, \chi_1)\big ) + 2 \epsilon^2\Re\big( \lambda_k \|\chi_1\|^2 + (\chi_1, \chi_0)\big) \,, 
\]
$(\phi, \chi_1) = 0 $ follows.  
In this way, $V_{\lambda_j} \perp V_{\lambda_k}$ is derived.  \hfill $\square$

\bigskip

Let us put  
\[
           V_0 = \bigoplus_{j: \Re \lambda_j = 0} V_{\lambda_j} \ \mbox{ and } \
           V_- = \bigoplus_{j: \Re \lambda_j <0 } V_{\lambda_j} \,.  
\]
Then, we have a orthogonal decomposition:
\[
         \C^{\n} = V_0 \bigoplus V_-   \,.
\]
Let $P_0$ be the orthogonal projection operator onto $V_0$.   
It is obvious that $[A, P_0] = O$.  
Now, let us put
\begin{equation}\label{A0}
                   A_0 = AP_0,  \quad A_- = A - A_0\,.
\end{equation}
Then
\[
            A_0 = A\upharpoonright _{V_0} \bigoplus O \,,  \ 
             A_- = O \bigoplus A\upharpoonright _{V_-}    \ \mbox{on}  \quad
              \C^{\n} = V_0 \bigoplus V_-   \,,  
\]
$ [A_0,A_-] =0$ and  $\displaystyle P_0 = \lim_{t \to \infty} e^{tA_-} $ hold.

Let $ \F_0 $ be a subspace of $\F$ spanned by vectors 
$ \{ (c, \xi_1) \cdots (c, \xi_n) |v\rangle  \, | \,  \xi_i \in V_0,  1 \ \leqslant i \leqslant n, \ n \geqslant 0 \, \} $.
Then the operator $\P_0$ on $\F$ characterized by 
\[
                \P_0(c, \xi_1) \cdots (c, \xi_n) |v\rangle 
                = (c, P_0\xi_1) \cdots (c, P_0\xi_n) |v\rangle     \qquad ( \xi_i \in \C^{\n}, \ 1\leqslant i \leqslant n)
\]
is the orthogonal projection onto $\F_0$.  

Let $\D_0$ be a subspace of $\D$ defined by
\[
               \D_0 = \{ \, \P_0\rho\P_0 \, | \, \rho \in \D \, \} \,. 
\]
We use a projection operator $\p$ from $\D$ to $\D_0$ which is defined by means of action on the basis (\ref{omegabase}) as 
\[
         \p\Phi( \xi_1, \cdots, \xi_n ; \eta_1, \cdots, \eta_m ) 
   = \Phi( P_0\xi_1, \cdots, P_0\xi_n ; P_0\eta_1, \cdots, P_0\eta_m ) \,. 
\]
From Proposition \ref{basisprop}(2) , we see that $\p$ and 
$\D \ni \rho \mapsto  \P_0\rho\P_0 \in \D_0$ are distinct projections.  
Proposition \ref{actA} implies the following lemma.  
\begin{lem}\label{proj}
\[
                   \lim_{t\to \infty}  e^{t\L(A_- , O)} \rho = \p\rho \,.   
\]
\end{lem}

\subsection{Asymptotic behavior} \label{AsmBeh}
In this subsection, we consider the asymptotic behavior of the solution of 
the master equation
\[
           \frac{d}{dt}\rho(t) = \L( A, M) \rho(t) 
\]
under the condition $ -A -A^{\dagger} \geqslant M \geqslant O$. 

\begin{thm}\label{AsyBe} 
\begin{equation}\label{AsyBe1}
               \rho(t) - e^{t\L( A_0, O)}e^{\L( O, M_{\infty})} \p\rho(0)  \rightarrow 0
      \quad \mbox{as } \ t \to \infty
\end{equation}
holds, where $M_{\infty} = \int_0^{\infty}e^{sA}Me^{sA^{\dagger}}ds$.  
Especially if $A_0 = O $, 
\begin{equation}\label{AsyBe2}
               \rho(t)  \rightarrow e^{\L( O, M_{\infty})} \Omega  
                \quad \mbox{as } \ t \to \infty
\end{equation}
holds.  
\end{thm}

{\sl Proof : } $1^{\circ} \ A_0M = MA_0 = O$ 

For $\phi \in V_{\lambda_j} \subset V_0$,  
$(\phi, (-A-A^{\dagger}) \phi) \geqslant (\phi, M\phi) \geqslant 0$ 
and $\Re \lambda_j = 0$ imply $(\phi, M \phi) =0$, hence 
$M\phi = M^{1/2}M^{1/2}\phi = 0 $ holds.   
So, we have $MP_0 = O $.   
Due to Hermeticity of $P_0$ and $M$, we also have $P_0M=O$.   
Together with (\ref{A0}), $ A_0M = MA_0 = O$ follows.  

$2^{\circ}$ Due to $1^{\circ}$, (\ref{cLAM}), $ A + A^{\dagger} = 0$ and $ [A_0, A_-] =o$, 
\[
          [ \L( A_0, O), \L( A_-, M) ] = \L( [A_0, A_-], A_0M + +MA_0^{\dagger}) = 0
\]
holds.  
From $1^{\circ}$ and th negativity of the real part of generalized eigenvalues of 
$A\upharpoonright _{V_-}$. we have
\[
   \int_0^t e^{sA_-} M e^{sA_-^{\dagger}}ds = \int_0^t e^{sA} M e^{sA^{\dagger}}ds
   \rightarrow M_{\infty}  \quad \mbox{as} \ t \to 0. 
\]
Therefore, 
\[
    \rho(t) = e^{t\L(A_0 + A_-, M)}\rho(0) = e^{t\L(A_0, O)} e^{t\L(A_-, M)} \rho(0)
\]
\[
            = e^{t\L(A_0, O)} e^{\L( O, \int_0^t e^{sA_-} M e^{sA_-^{\dagger}}ds) }
            e^{t\L(A_-, O)} \rho(0)
\]
holds,  
Together with Lemma \ref{proj} and the continuity of $ (A, M) \mapsto \L(A,M)$, we get (\ref{AsyBe1}).  

$3^{\circ}$ If $A_0= O$, $ V_0 = \{ 0 \}$ and $P_0 = O$ hold. 
So, $\D_0$ is one dimensional subspace spaned by $\Omega$ and (\ref{AsyBe2}) follows from (\ref{AsyBe1}).    \hfill $\square$

\bigskip

Posibility of synchronized asymptotic behavior is showin in (\ref{AsyBe1}) for 
$V_0$ having small but non-zero dimensions.\cite{GCZ}    
On the other hand, (\ref{AsyBe2}) shows a relaxation phenomenon to a unique steady state  if dim$\, V_0 = 0$.       
\section{Applications}
In this section, a few applications of them are presented to show how to use the formalism stated so far.  
\subsection{Gauss states} 
The steady states for the quadratic Fermion systems are known to be Gaussian states.  
In this section, we derive some basic properties of Gaussian states from our view point.  
We also see the correspondence between the time evolution of the fermionic Gibbs states 
in terms of the master equations and affine transformation on $\M( \n, \C)$.  
We begin with the following Lemma.  

\begin{lem}\label{GL}
For Hermite $M, T \in \M( \n. \C) $,  
\begin{equation}\label{gl}
        e^{t\L(-M/2, M)} \hG(T) = \hG(e^{tM/2}Te^{tM/2})e^{t\L(-M/2, M)}
\end{equation}
holds.  
\end{lem} 
{\sl Proof : } Because of
\begin{equation}\label{811}
        \L( -M/2, M) = \hG(M) + \hF(M/2) + \hB(M/2) - \tr M
\end{equation}
 (\ref{LAM}) and Lemma \ref{lem1}, we obtain   
\begin{equation}\label{LGG}
 [ \L( -M/2, M), \hG(T)] = \hG\Big(\frac{MT + TM}{2}\Big) \,.
 \end{equation} 
Put 
\[
                    g(s) =  e^{s\L(-M/2, M)} \hG(e^{(t-s)M/2}Te^{(t-s)M/2})
                    e^{(t-s)\L(-M/2, M)}  \,.  
\]
Then by the above commutation relation, we get $g'(s) = 0$ and accordingly $g(t) = g(0)$ which implies the claim.    \hfill $\square$ 

\begin{prop}\label{gauss}
Let $R, T \in \M( \n, \C)$ be Hermite. 
We assume that all eigenvalues of $R$ are contained in $(0, 1)$.   
Then
\begin{equation} \label{Gauss}
           e^{\L( O, R)}\Omega  = \det ( I - R) e^{ (c, \log(R(I-R)^{-1}) c)}  
\end{equation}
and
\begin{equation}\label{expGauss}
         \Tr\big[ (c, Tc)e^{\L( O, R)}\Omega\big] = \tr(TR) 
\end{equation}
hold.  
\end{prop}
{\sl Proof : } From Theorem \ref{thmTimeEvol} and the fact $\L(A, O)\Omega = 0$, we have
\begin{equation}\label{LOmega}
e^{L( -M/2, M)}\Omega = e^{L( O, \, I -e^{-tM})}\Omega 
\end{equation}
for Hermite $M \in \M( \n, \C)$.  

On the other hand, due to (\ref{811}) and (\ref{gl}) 
\[
            \frac{d}{ds}e^{s\L(-M/2, M)}\Omega = e^{s\L(-M/2, M)} (\hG(M)- \tr M)\Omega 
            = (\hG(Me^{sM}) - \tr M)e^{s\L(-M/2, M)}\Omega 
\]
holds.   
So, for  
\[
          h(s) = e^{-(t-s)\tr M}e^{\hG(e^{tM} - e^{sM})} e^{s\L( -M/2, M)}\Omega \,.
\]
we have $d h(s)/ds = 0$.   
It follows that  $h(t) = h(0)$, i.e., 
\[
             e^{t\L(-M/2, M)}\Omega = e^{\hG(e^{tM} - I ) - t\tr M}\Omega \,.   
\] 
Together with (\ref{LOmega}), 
\begin{equation}\label{LG}
           e^{L( O, \, I -e^{-tM})}\Omega  = e^{\hG(e^{tM} - I) - t\tr M}\Omega 
\end{equation}
is derived.  

For Hermite $X\in \M( \n, \C)$ whose eigenvalues are all positive, 
we put $\xi_j \,, x_j \ ( j = 1, \cdots \n)$ for its normalized eigenvectors and corresponding eigenvalues.     
Then, $X = \sum_j x_j\xi_j \xi_j^{\dagger}$ and therefore 
\[
      e^{\hG(X)}\Omega = \big[\prod_je^{x_j\hG(\xi_j\xi_j^{\dagger})} \big]\Omega 
      = \big[\prod_j(1 + x_j\hG(\xi_j\xi_j^{\dagger}))\big]\Omega 
      = \sum_{\nu \in \{0,1\}^{\n}} x^{\nu} (c, \xi)^{\nu} \Omega\{(c, \xi)^{\nu}\}^{\dagger}  
\]
hold, where the multi-index notations (\ref{multiindex}) are used.   

The expression shows that the operator $e^{\hG(X)}\Omega$ on $\F$ has eigenvector
$(c, \xi)^{\nu} |v\rangle $ with corresponding eigenvalue $x^{\nu}$ for each 
$\nu \in \{0 ,1\}^{\n}$, since 
$\langle v|\{(c, \xi)^{\nu}\}^{\dagger} (c, \xi)^{\mu}|v \rangle =\delta_{\mu,\nu} $.      
On the other hand, $\sum_j \log x_j (c, \xi_j)(\xi_j, c)$ has eigenvector
$(c, \xi)^{\nu} |v\rangle $ with corresponding eigenvalue $\sum_j \nu_j\log x_j$ for each $\nu \in \{0 ,1\}^{\n}$.  
Hence, we get
\[
            e^{\hG(X)}\Omega = e^{\sum \log x_j (c, \xi_j)(\xi_j, c)} = e^{(c, (\log X) c)} \,. 
\]
By setting $ R= I -e^{-tM} , e^{tM} - I  = X$, this equation and (\ref{LG}) leads to (\ref{Gauss}).   

For (\ref{expGauss}), we note that
\[
             \hG(T) e^{\L( O, I-e^{-tM})}\Omega = \hG(T) e^{t\L( -M/2, M)}\Omega
              =e^{t\L( -M/2, M)} \hG(e^{-tM/2} T e^{-tM/2})\Omega  \,.  
\]
holds because of (\ref{LOmega}) and Lemma \ref{GL}.   
Recalling that $e^{t\L( -M/2, M)} $ is trace preserving, we get   
\begin{equation}\label{T=T}
              \Tr[\hG(T) e^{\L( O, I-e^{-tM})}\Omega] 
                = \Tr[\hG(e^{-tM/2} T e^{-tM/2})\Omega] \,.  
\end{equation}
On the other hand the identity
\[
            \Tr[\hG(T)\rho] = \Tr[ (\tr T - (c, Tc))\rho]
\]
is derived from  Definition \ref{def1} and trace formula.  
We apply this to both side of (\ref{T=T}) to get 
\[
            \tr T - \Tr[(c, Tc) e^{\L( O, I-e^{-tM})}\Omega] = \tr(Te^{-tM}) \,,    
\]
where we have used  Definition \ref{def1} and  trace preserving property 
of $e^{\L( O, I-e^{-tM})}$ and trace formula.  
By putting $ R = I - e^{tM} $, (\ref{expGauss}) follows.   \hfill $\square$   

\begin{rem}\label{entropy}
Combining (\ref{Gauss}) and (\ref{expGauss}), an expression of the Gibbs state is expressed as
\[
                \mathrm{Ent}[e^{\L( 0, R)}\Omega ] = -\tr(R\log R) - \tr((I-R)\log (I-R)) \,. 
\]
\end{rem}

The following proposition concerns time evolution of Gibbs states.  

\begin{prop}\label{evolutionGauss}
For $A \in \M( \n, \C)$ and Hermite $M, R \in \M( \n, \C)$, 
\begin{equation}
   e^{t\L( A, M)} e^{\L( O, R)}\Omega = e^{\L( O, R(t))} \Omega
\end{equation}
holds, where $ R(t) = e^{tA}Re^{tA^{\dagger}} +  \int_0^te^{sA}Me^{sA^{\dagger}}ds$.
\end{prop}
{\sl Proof : } Due to Theorem \ref{thmTimeEvol} and Lemma \ref{com2}, 
we obtain
\[
   e^{t\L( A, M)} e^{\L( O, R)}\Omega 
  = e^{\L( O, \, \int_0^te^{sA}Me^{sA^{\dagger}}ds)} 
       e^{t\L( A, O)} e^{\L( O, R)}\Omega  
\]
\[
   = e^{\L( O, \, \int_0^te^{sA}Me^{sA^{\dagger}}ds)}
     e^{\L( O, \, e^{tA}Re^{tA^{\dagger}})}e^{t\L( A, O)}\Omega \,.  
\]
Because of $[\L( O, N_1), \L( O, N_2)] = 0$ and 
$\L( A, O) \Omega = 0$, the claim follows.  \hfill $\square$ 

\begin{rem}
From Lemma \ref{formulae1} (1) and (\ref{AT}), 
\[
       R(t) = e^{t(\!( A, M)\!) } R
\]
holds. 
It shows that the time evolution of Gibbs states is described by the 
affine transformation on Hermete matrices in $\M( \n, \C)$.   
\end{rem}

\subsection{Skin effect}
Here we consider how to incorporate non-Hermite quantum mechanical model into 
the GKSLD theory for open system, taking  the Hatano-Nelson model as an example.   
As the non-Hermete Hamiltonian corresponding to the open system controlled by Liouvillian (\ref{Liou1}), we asign  
\begin{equation}\label{nHH}
                   H_{\mathrm{nH}} = H - iD + i E = iA +iM \,. 
\end{equation}

For the Hatano-Nelson  model as one-dimensional lattice model, we set   
\[
       H = \omega I + \lambda \mathbb{F}, \qquad 
       D- E = \gamma(aI + \mathbb{G})  \,, 
\]
where 
\[
     {\mathbb F}_{j,k} = \delta_{j, k+1} + \delta_{j+1,k} \,, \qquad 
     {\mathbb G}_{j,k} = i\delta_{j, k+1} -i \delta_{j+1,k}
\]       
Thus, $H_{\mathrm{nH}} = (\omega -ia\gamma)I + \mathbb{K}$, 
where 
\[
      {\mathbb K}_{j,k} = (\gamma + \lambda)\delta_{j, k+1} 
                 - (\gamma - \lambda) \delta_{j+1,k}  \,. 
\]
For definiteness, let us consider the case 
$\omega >0, \ \gamma>\lambda >0, \ a > 2$.  
Set $V(\kappa) \in \M( \n; \C)$ as
\[
      V(\kappa)_{jk} = \kappa^{j-1}\delta_{jk}  \quad \mbox{for} \quad  
      \kappa = \sqrt{\frac{\gamma - \lambda}{\gamma + \lambda}} < 1 \,. 
\]
Then,
\[
            V(\kappa)H_{\mathrm{nH}} V(\kappa)^{-1}  =
             (\omega -ia\gamma)I - i \sqrt{\gamma^2 -\lambda^2}
       \,\mathbb{G}  
\]
is Hermite, and it shows that the eigenvectors of $H_{\mathrm{nH}}$
have negligibly small components away from the bottom (skin effect).  

As indicated in \cite{MHC}, the choice of $E$ and $D$ is crucial 
to nature of $X$.   
We try to find a suitable $E$ such that the asymptotic steady state 
$ e^{\L(O, X)}\Omega \,$ for 
\begin{equation}\label{X}
      X = \int_0^{\infty}e^{sA}Me^{sA^{\dagger}} ds 
             =  \int_0^{\infty}e^{s(-iH_{\mathrm{nH}} -2E)}2E
               e^{s(-iH_{\mathrm{nH}} -2E)^{\dagger}} ds 
\end{equation}
have a property which exhibit ``skin effect".  
Our answer is : 

\begin{thm}\label{skineffect}
If we set
\[
        E =  \int_0^{\infty} e^{(2X - I)s}xV(\kappa)^{-1}
        [2a\gamma I + 2\sqrt{\gamma^2 - \lambda^2}\mathbb{G}]
         V(\kappa)^{-1}e^{(2X - I)s} ds \,, 
\]
then $M = 2E$ and $A = -iH_{\mathrm{nH}} - M$ satisfy the condition 
$ -A - A^{\dagger} \geqslant M \geqslant O$ and 
$\displaystyle      X= xV(\kappa)^{-2}  $,  
where we may choose $x \in ( 0, \kappa^{2\n -2}/2)$.  
\end{thm}
{\sl Proof : } Due to (\ref{X}), $X$ satisfies
\[
         (-iH_{\mathrm{nH}} -2E)X + X( -iH_{\mathrm{nH}} -2E)^{\dagger}
          = \int_0^{\infty}\frac{d}{ds}\big(e^{s(-iH_{\mathrm{nH}} -2E)}2E
               e^{s(-iH_{\mathrm{nH}} -2E)^{\dagger}} \big)ds = -2E,   
\]
where we have noted that the condition $a >2$ makes the integral convergent.   
Hence we have
\[
            E(2X-I) + (2X-I)E = -i(H_{\mathrm{nH}} X - XH_{\mathrm{nH}}^{\dagger}) \,, 
\]
from which we get the expreshon 
\[
          E = \int_0^{\infty} e^{(2X-I)s} i(H_{\mathrm{nH}} X - XH_{\mathrm{nH}}^{\dagger})
                e^{(2X-I)s} \,ds \,,  
\]
since $X = xV(\kappa)^{-2} $ and the condition $x \in ( 0, \kappa^{2\n -2}/2)$ makes the integral convergent.  
We also have 
\[ 
           i(H_{\mathrm{nH}} X - XH_{\mathrm{nH}}^{\dagger}) = xV(\kappa)^{-1}
           [2a\gamma I + 2\sqrt{\gamma^2 -\lambda^2} \mathbb{G}]V(\kappa)^{-1}  
           \geqslant O \,.   
\tag*{$\square$}
\] 
\begin{rem}
\begin{description}
\item[\rm{(1)}] 
Because of $X_{jk} =x\kappa^{2- 2j}\delta_{jk}$, we see that the components of $X$ 
away from the bottom are exponentially small.  
Due to (\ref{expGauss}), it means that $e^{\L( O, X)}\Omega$ is a state of concentration of fermions to one end of the 1-dimensional lattice.   
\item[\rm{(2)}] 
By the ``simplest" choice 
\[
       E = \delta D = \frac{i\delta}{2(1-\delta)}
  (H_{\mathrm{nH}} - H_{\mathrm{nH}}^{\dagger}) \,,
\]
the steady state which has no characteristic feature 
of skin effect is derived.  
In fact, due to (\ref{X}), we have
\[
       X = -\frac{\delta}{1+\delta} \int_0^{\infty}\frac{d}{ds}
       \big(e^{-iH_{\mathrm{nH}}-2E} 
        e^{(-i H_{\mathrm{nH}} -2E)^{\dagger}}\big)ds  
       = \frac{\delta}{1+\delta} I
\]
holds for general $ H_{\mathrm{nH}} $ (which lets the above integral be  
convergent).   
For other answers and speculations, we refer to \cite{MHC}.  
\end{description}
\end{rem}

\section{Toward possible generalizations}
In this section, we glance at other quadratic interacting systems to investigate whether 
our method is generalizable.   
Commutation relations among Liouvillians of these respective systems which are related to certain affine transformations are given.     
\subsection{General quadratic Fermion systems}
We use the following Majorana operators:
\[
   w_{2m-1} = c_m + c_m^{\dagger}\,, \quad w_{2m} = i(c_m - c_m^{\dagger})\,,  
    \quad \{ w_j, w_k \} = 2\delta_{jk}  \quad (j, k = 1, \cdots, 2\n) \,. 
\]
to express $\n$ Fermion systems.    
Then, the Liouvillian for the general  quadratic $\n$ Fermion systems is written by 
\begin{equation}\label{gFL}
          \L\rho = \sum_{j,k =1}^{2\n}\big( -iH_{jk}[w_jw_k, \rho] 
          + M_{jk}(2w_j\rho w_k - \{w_kw_j, \rho \} ) \big) \,, 
\end{equation}
where $H$ is Hermite $2\n \times2 \n$ matrix.   
Without loss of generality, it is assumed to be anti-symmetric.  
Hence, $H$ consists of pure imaginary components.  
$M$ is a Hermete positive $2\n \times2\n$ matrix.\cite{P}

Let us define operator $\L( A, N) $ acting on $\D$ for $2\n \times2\n$ real matrix
$A$ and $2\n \times2\n$ real anti symmetric matrix $N$ by
\begin{align*}
             \L( A, N)\rho &= \frac{1}{4}\sum_{j,k =1}^{2\n}
             \Big( \frac{(A- {}^t\!A)_{jk}}{2}[w_jw_k, \rho] 
\\
              & +iN_{jk}  \{w_jw_k, \rho \}  + (-A - {}^t\!A +2iN)_{jk}w_j\rho w_k \Big) \,.     
\end{align*}
Under the additional condition $-A - {}^t\!A + 2iN \geqslant O$, 
$H = (A^{\dagger} -A)/2i $ and $ M = iN -(A+A^{\dagger})/2$ satisfy the 
above condition for $A$ and $N$, and 
\[
    4\L(A, N) + 2\tr A = 4\L(-iH - (M + {}^t\!M)/2,  (M - {}^t\!M)/2i) - 2\tr M
\]
reproduce the Liouvillian $\L$ in (\ref{gFL}).    

The commutation relation
\[
     [\L(A, N), \L(B, R)] = \L([A, B], AR + R{}^t\!A - BN - N{}^t\! B) 
\]
holds for $A, B, N, R \in \M(\n, \R)$ satisfying $N + {}^t\!N = R + {}^t\!R = O$.    

Although we omit detailed derivation, it would be confirmed by straightforward calculation.   
\subsection{Gauge invariant quadratic Boson systems}
Let  $a_j, \ a^{\dagger}_j \ ( j= 1, \cdots, \n) $ be the 
annihilation and the creation operators defined in the
Fock space $\mathscr{F}_B$ generated by a cyclic vector 
$| v \rangle $.
That is, the Hilbert space $\mathscr{F}$ is the completion of 
the algebraic span $\mathscr{F}_{B\mbox{\tiny fin}}$ of vectors 
\[
  \{ \, a_1^{\dagger \nu_1}\cdots a_{\n}^{\dagger\nu_{\n}}
| v \rangle \, | \, \nu_1, \cdots, \nu_{\n} = 0, 1, 2, \cdots \, \}
\]
and $a_j, a^{\dagger}_k$ satisfy the Canonical Commutation 
Relations (CCR)
\begin{equation}\label{CCR0}
      [a_j, a^{\dagger}_k] = \delta_{jk}, 
      \quad  [a_j, a_k] = 0, \quad 
         [a_j^{\dagger}, a_k^{\dagger}] = 0
      \quad \mbox{on} \quad \mathscr{F}_{\mbox{\tiny fin}} \,.
\end{equation}

We deal with the Liouvillians for the gauge invariant quadratic 
$\n$-Boson systems of GKSLD type
\[
     \L\rho(t) = -i[\mathcal{H}, \rho(t)] 
         + \sum_{m=1}^{S} \big( 2\mathcal{D}_m\rho(t)  \mathcal{D}^{\dagger}_m 
        - \{\,  \mathcal{D}^{\dagger}_m  \mathcal{D}_m, \rho(t) \, \} \big) 
\]
\begin{equation}
        + \sum_{m=S+1}^{S+S'} \big( 2\mathcal{D}^{\dagger}_m\rho(t)  \mathcal{D}_m 
        - \{\,  \mathcal{D}_m  \mathcal{D}^{\dagger}_m, \rho(t) \, \} \big) 
\end{equation}
defined at present on the space of finite rank operators on $\F_{B\mbox{\tiny fin}}$; 
\[
                \mathfrak{B}(\F_{\mbox{\tiny fin}}) = 
              \mbox{linear span of } 
              \big\{ \, a_1^{\dagger\nu_1}\cdots a_{\n}^{\dagger\nu_{\n}}
          | v \rangle \langle v | 
       a_{\n}^{\mu_{\n}}\cdots a_1^{\mu_1}
        \, \big| \, \nu_1, \cdots, \nu_{\n}, \mu_1, \cdots, \mu_{\n} = 0, 1, 2, \cdots \, \big\} \,. 
\]
Here, 
\[
    \mathcal{H} = \sum_{j, k = 1}^{\n} H_{jk} a^{\dagger}_ja_k , \quad
    \mathcal{D}_m = \sum_{j=1}^{\n}\overline{\ell_{mj}}a_j = (\ell_m, a) 
\]
with Hermite $H \in \M(\n; \C)$ and $\ell_m \in \C^{\n}$.    

Now, let us put
\begin{align}
               A = -iH -\sum_{m=1}^S\ell_m\ell_m^{\dagger} + 
               \sum_{m=S+1}^{S+S'}\ell_m\ell_m^{\dagger} \,,
\notag \\
               M = \sum_{m=1}^S\ell_m\ell_m^{\dagger} + 
               \sum_{m=S+1}^{S+S'}\ell_m\ell_m^{\dagger} \,. 
\label{gaubAM}
\end{align}
Then, $A, M \in \M(\n ; \C)$ satisfy the condition  
$ -2M \leqslant A + A^{\dagger} \leqslant 2M $. 
Contrary, the condition reproduces (\ref{gaubAM}).  

Now we define $\L( A, M)$ by  
\[
    \L( A, M) \rho = \sum_{j,k =1}^{\n}\Big( 
    \frac{1}{2}(A - A^{\dagger})_{jk}[a_j^{\dagger} a_k, \rho]  
   + \frac{1}{2}( A +A^{\dagger})_{jk}(a_j^{\dagger}\rho a_k - a_k\rho a_j^{\dagger})
\]
\begin{equation}
                 -M_{jk}[a_j^{\dagger}, [a_k, \rho]] \, \Big)  
\label{gaubLAM}
\end{equation}
such that
\[
        \L( A, M) = \L + \sum_{m=1}^S( \ell_m, \ell_m) - 
                  \sum_{m=S+1}^{S+S'}(\ell_m, \ell_m)   
\]
holds.   

We extend (\ref{gaubLAM}) to general $A, M \in \M(\n, \C)$.  
Then, the commutation relation 
\[
      [\L(A, M), \L(B, N)] = \L([A, B], AN + NA^{\dagger} - BM- MB^{\dagger}) 
\]
holds on $ \mathfrak{B}(\F_{B\mbox{\tiny fin}}) $ for 
$A, B, M, N \in \M(\n, \C)$.     

\subsection{General quadratic Boson systems}  
We put $b_j = a_j. \ b_{\n+j} = a_j^{\dagger} $. 
And we consider the Liouvillian 
\[
     \L\rho(t) = -i[\mathcal{H}, \rho(t)] 
         + \sum_{m=1}^{S} \big( 2\mathcal{D}_m\rho(t)  \mathcal{D}^{\dagger}_m 
        - \{\,  \mathcal{D}^{\dagger}_m  \mathcal{D}_m, \rho(t) \, \} \big) 
\]
for $\rho \in \mathfrak{B}(\F_{B\mbox{\tiny fin}})$, where
\[
    \mathcal{H} = \sum_{j, k = 1}^{2\n} H_{jk} b_j b_k ,
\]
and
\[
    \mathcal{D}_m = \sum_{j=1}^{2\n}\phi_{mj} b_j \,.
\]
Now we put
\[
   A = (-2iH + D - {}^t\!D)\tau\sigma \,, 
       \quad M =-(D + {}^tD)\tau\sigma \,, 
\]
where          
\[
       D_{jk} = \sum_{m=1}^{\tilde{M}} \phi_{mj}(\phi_m^{\dagger}\sigma)_k  \,,
          \quad
      \sigma = \begin{pmatrix}
                        O & I \\
                        I & O                    
               \end{pmatrix} \,, \quad 
      \tau = \begin{pmatrix}
                        I & O \\
                        O & -I                    
               \end{pmatrix} \ \in \M( 2\n; \C) \,.  
\]
Note that CCR  imply $[b_j, b_k] = (\tau\sigma)_{jk}$.  

The condition $D\sigma \geqslant O $ holds and without loss of generality we may assume ${}^t\!H = H$.  
We also assume  $H^{\dagger} = \sigma H\sigma$ to make $\mathcal{H}$ symmetric, 
i.e., $\mathcal{H}^{\dagger} \supset \mathcal{H}$.    
The set of these conditions on $H$ and $D$ is equivalent to the set of conditions on $A$ and $M$
\[
         \tau(A-A^{\flat})^{\dagger} +(A-A^{\flat})\tau = O \,, 
        \ (A + A^{\flat} + 2M)\tau \geqslant O \,, \ M + M^{\flat} = O \,,  
\]
where we have used the involution 
$A^{\flat} = \sigma \tau^t\!A\tau\sigma$.

Let us put
\begin{equation}
       \L( A, M)\rho = \frac{1}{2}\sum_{j,k =1}^{2\n}(A\sigma\tau)_{jk} \big(
       [b_jb_k, \rho] + b_j\rho b_k - b_k\rho b_j \big) 
       + \frac{1}{2}\sum_{j,k =1}^{2\n}(M\sigma\tau)_{jk}[b_j, [b_k, \rho]]   
\label{genbLAM}
\end{equation}
to make the equality 
\[
              \L = \L(A,M) + \tr(D\sigma\tau)
\]
holds.

Let us extend (\ref{genbLAM}) to general $A, M \in \M( 2\n; \C)$ satisfying condition $ M + M^{\flat} = O$.   
Then the commutation relation 
\[
          [\L(A, M), \L(B, N)] = \L([A, B], AN + NA^{\flat} - BM- MB^{\flat}) 
\]
holds on $\mathfrak{B}(\F_{B\mbox{\tiny fin}})$ for 
$A, B, M, N \in \M( 2\n; \C)$ with condition 
$ M + M^{\flat} = N + N^{\flat} = O$.     

\bigskip

\[
\mbox{\textbf{\large Acknowledgments}}
\]

\noindent
This work was supported by JSPS KAKENHI Grant Number JP17K05272.  
The author is grateful to Professor Valentin A. 
Zagrebnov and Professor Hajime Moriya for useful discussions and 
suggestions.  

\bigskip


\end{document}